\documentclass[reprint,superscriptaddress,amssymb,amsmath,aps,showpacs,10pt,longbibliography,prb]{revtex4-2}

\usepackage{graphicx}
\pdfoutput=1

\graphicspath{{figures/}}

\def\jjj{$J_1$--$J_1'$--$J_2$}

\usepackage{color}

\definecolor{darkred}{rgb}{0.45,0.02,0} 

\definecolor{darkgreen}{rgb}{0.02,0.45,0.0}
\definecolor{darkblue}{rgb}{0,0.02,0.45}

\usepackage[colorlinks,bookmarks=false,citecolor=darkblue,linkcolor=red,urlcolor=blue]{hyperref}
\usepackage{multirow}

\def\be{\begin{equation}}
\def\ee{\end{equation}}
\def\bea{\begin{eqnarray}}
\def\eea{\end{eqnarray}}
\newcommand{\ra}[1]{\renewcommand{\arraystretch}{#1}}

\def\mc{\mathcal}

\usepackage{booktabs}

\usepackage{verbatim}

\begin{document}


\title{One-dimensional physics of the frustrated quantum magnet PHCC}

\author{Alexander A. Tsirlin}
\email{altsirlin@gmail.com}
\affiliation{Felix Bloch Institute for Solid-State Physics, University of Leipzig, 04103 Leipzig, Germany}

\author{Oleg Janson}
\email{o.janson@ifw-dresden.de}
\affiliation{Institute for Theoretical Solid State Physics, Leibniz IFW Dresden, 01069 Dresden, Germany}

\author{Ioannis Rousochatzakis}
\email{i.rousochatzakis@lboro.ac.uk}
\affiliation{Department of Physics, Loughborough University, Loughborough LE11 3TU, United Kingdom}


\begin{abstract}
We report a comprehensive microscopic study of the frustrated quantum magnet PHCC, (C$_4$H$_{12}$N$_2$)Cu$_2$Cl$_6$, using density-functional band-structure calculations combined with numerical quantum many-body simulations of the underlying spin Hamiltonian. We show that the magnetism of PHCC is captured by a one-dimensional model of the frustrated spin chain with alternating nearest-neighbor couplings ($J_1=23.1$\,K, $J_1'=7.0$\,K) and uniform next-nearest-neighbor couplings ($J_2=13.9$\,K). This model, which can also be thought of as a zigzag ladder, provides a quantitative description of the magnetic susceptibility and the magnetization process, and accounts for the observed dispersion of the single-triplet band and its merging into a continuum near the Brillouin zone center. We also make predictions for the existence of sharp bound (anti-bound) states of two triplets, below (above) the bottom (upper) edge of the two-particle scattering continuum.
\end{abstract}

\maketitle


\section{Introduction}
The field of one-dimensional (1D) magnetism has been a fertile playground for exploring novel quantum states of matter for many decades~\cite{qmagnetism}. Exact solutions of certain 1D spin models have played a pivotal role in understanding a wide range of many-body phenomena, from quantum and thermal phase transitions to spin liquids, fractionalization and emergent fermions~\cite{Ising1925,JordanWigner1928,Bethe1931,Onsager1944,Schulz1964,Mikeska2004}. A key advantage of 1D spin models is that they are amenable to a wide range of theoretical approaches (most notably, Bethe ansatz, mapping to fermions and bosons, conformal invariance, nonlinear $\sigma$-models, and series expansions), but also to extremely accurate numerical quantum many-body methods (from density-matrix renormalization group (DMRG) and matrix product states to quantum Monte Carlo). At the same time, the field has been driven by the realisation of 1D spin models in real-world candidate materials. For example, the rich physics of the 1D Heisenberg chain and the two-leg ladder have been extensively studied in both theory~\cite{qmagnetism} and experiment~\cite{mourigal2013,ruegg2008,thielemann2009,ward2017,blosser2018,schmidiger2012,schmidiger2013}. 

Material realizations of more complex 1D spin lattices have been reported as well. However, the increasing complexity of their crystal structures and underlying magnetic interactions make mapping onto exact (idealized) spin models ambiguous~\cite{jeschke2011}, and even magnetic dimensionality is sometimes vividly debated in the literature~\cite{janson2010,janson2016}.

Here, we report a microscopic study of the frustrated quantum magnet PHCC (piperazinium hexachlorodicuprate) with the chemical formula (C$_4$H$_{12}$N$_2$)Cu$_2$Cl$_6$~\cite{daoud1986,battaglia1988}. This material -- featuring $S=\frac12$ spins localized on Cu$^{2+}$ sites -- is one of the early candidates for the experimental realization of a quantum spin liquid. It features multiple frustrated magnetic interactions and a gapped singlet ground state without any signatures of a symmetry-breaking magnetic transition~\cite{daoud1986}, although magnetic order is known to set in under applied field~\cite{stone2006b,stone2007} or upon application of pressure~\cite{hong2010,thede2014,perren2015,bettler2017} via closing of the spin gap. The magnetic excitations of PHCC have been extensively studied by inelastic neutron scattering. The low-energy part of the spectrum is dominated by a coherent triplet band that merges into a continuum at higher energies~\cite{stone2006}. These features manifest two different facets of the material. While the triplet band is typical for gapped quantum magnets with weakly interacting spin dimers~\cite{zapf2014}, the continuum is reminiscent of a uniform spin chain with spinon excitations~\cite{mourigal2013}, or of a more complex frustrated magnet with the (proximate) spin-liquid ground state where microscopic descriptions in terms of either spinons or interacting magnons can be envisaged~\cite{zheng2006,chernyshev2009}. Moreover, the triplet band in PHCC merges into the continuum and manifests a model case for the quasiparticle breakdown~\cite{stone2006}, a fingerprint of the underlying many-body state reported in various interacting quantum systems ranging from triangular antiferromagnets to superfluid helium~\cite{verresen2019}.

Despite the extensive experimental coverage, only a limited theoretical analysis of PHCC has been reported to date. A complex two-dimensional spin lattice with multiple frustrated interactions was inferred from the dispersion of the triplet band~\cite{stone2001}, but the values of individual exchange parameters could not be determined. Here, we revise this magnetic model and propose a simpler and microscopically justified description in terms of a 1D spin model that captures thermodynamic properties as well as the main features of the excitation spectrum. We demonstrate that, despite its complex structure, PHCC belongs to the group of quasi-1D magnets and manifests a rare example of the frustrated spin chain with alternating nearest-neighbor couplings $J_1$ and $J_1'$ and a second-neighbour coupling $J_2$. This model, which can also be viewed as a zigzag spin ladder, has been discussed in the context of the spin-Peierls material CuGeO$_3$~\cite{Uhrig1996,Uhrig1997}, and has attracted a lot of attention from theory~\cite{Brehmer1996,Pati1997,Brehmer1998,Totsuka1998,Mila1998,Bouzerar1998,Singh1999,Kotov1999b,Mori1999,Muller2000,Mikeska2004}.


\section{Methods}
Magnetic couplings in PHCC were obtained from density-functional-theory (DFT) band-structure calculations performed in the \texttt{FPLO}~\cite{fplo} and \texttt{VASP}~\cite{vasp1,vasp2} codes using the Perdew-Burke-Ernzerhof (PBE) version of the exchange-correlation potential~\cite{pbe96}. The lattice parameters and atomic positions for Cu, Cl, N, and C were taken from Ref.~\cite{daoud1986}, whereas hydrogen positions were relaxed because of possible errors due to the lower sensitivity of x-rays to hydrogen atoms. Different settings of the triclinic PHCC structure are reported in the literature. Here, we adopt the setting of Ref.~\cite{battaglia1988}, which is consistent with the later neutron-scattering studies. This setting requires the swap of the $\mathbf a$ and $\mathbf c$ axes in the structure from Ref.~\cite{daoud1986}.

Magnetization~\cite{stone2007} and electron-spin-resonance~\cite{glazkov2012} measurements on PHCC single crystals indicate only a weak magnetic anisotropy. Therefore, we restrict ourselves to the isotropic spin model,
\be
\mc{H}=\sum_{\langle ij\rangle}J_{ij} {\bf S}_i\cdot{\bf S}_j\,,
\ee
where $\mathbf S_j$ and $\mathbf S_j$ are $S=\frac12$ operators on sites $i$ and $j$, and summation is performed over bonds $\langle ij\rangle$. The exchange couplings $J_{ij}$ are obtained by a mapping procedure~\cite{xiang2011,tsirlin2014} using total energies of collinear spin configurations calculated on the DFT+$U$ level. This incorporates the on-site Coulomb repulsion parameter $U_d$, the Hund coupling $J_d$ applied to the correlated $3d$ shell of the Cu atoms, and the double-counting correction in the atomic limit. Several supercells were used to resolve all couplings with the Cu--Cu distances below 10\,\r A. Additionally, we used the tight-binding analysis of the nonmagnetic PBE band structure to determine the underlying hoppings and visualize magnetic orbitals via the Wannier functions implemented in \texttt{FPLO}~\cite{koepernik2023}.

The magnetic susceptibility $\chi$ of the infinite {\jjj} chain model was simulated using the transfer-matrix density-matrix renormalization-group (TMRG) approach~\cite{wang97}. We kept 160 states during truncation and the maximal Trotter number of $1.2\times10^4$. The starting inverse temperature $\beta$ was set to 0.1$J_1$. Since the accessible temperatures are $\beta\left(\frac{1+n}{2}\right)$, where $n\in\mathbb{N}$, the sampling becomes very sparse at high temperatures. To improve it, we additionally solved the spin Hamiltonian on 16-site chains with periodic boundary conditions by exact diagonalization (ED), as implemented in \texttt{ALPS} version 2.3~\cite{alps1.3, alps2.0}. By comparing the two data sets, we find that above $t^*\!=\!T/J_1\!=\!1$  (where $J_1$ is the dominant exchange coupling, see below) the TMRG- and ED-calculated susceptibilities $\chi^*$ are practically identical (deviations are smaller than $10^{-6}$). This allows us to combine TMRG data for $\chi(t^*\!\leq{}\!1)$ and ED data for $\chi(t^*\!>\!1)$ magnetic susceptibilities into a single data set. Magnetization curves of $128$-site chains at zero temperature are simulated using density-matrix renormalization group (DMRG) code from \texttt{ALPS} version 2.3~\cite{alps1.3, alps2.0}. The maximum number of kept states is chosen to be 200, which led to a reasonable truncation error of $\lesssim{}2\times{}10^{-6}$. Excitation spectra of the {\jjj} model at zero temperature are calculated using sparse ED~\cite{alps1.3,alps2.0} on finite lattices of 24 sites. In all ED and DMRG simulations we use periodic boundary conditions.


\section{Microscopic magnetic model}
\begin{figure}
\includegraphics{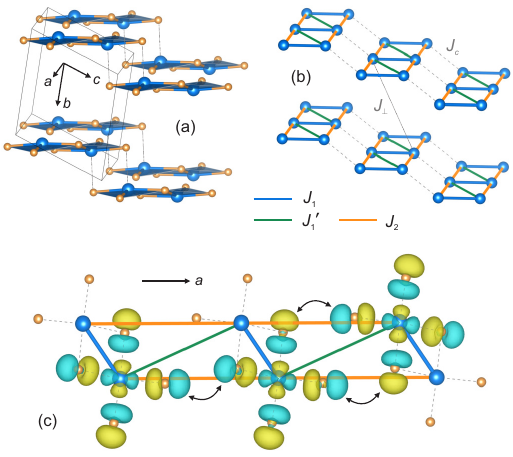}
\caption{\label{fig:structure}
(a) Crystal structure and (b) magnetic model of PHCC. Piperazinium ions are omitted for clarity. The dotted lines in the left panel show the long Cu--Cl bonds of 2.606\,\r A that are perpendicular to the CuCl$_4$ plaquettes. (c) The {\jjj} frustrated zigzag chain with the relevant orbitals visualized using Wannier functions composed of the Cu $d_{x^2-y^2}$ and Cl $3p$ states. The arrows indicate the overlap of the adjacent Cl $3p$ orbitals resulting in the extended Cu--Cl--Cl--Cu superexchange pathways.
}
\end{figure}

\begin{figure}
\includegraphics[width=8.5cm]{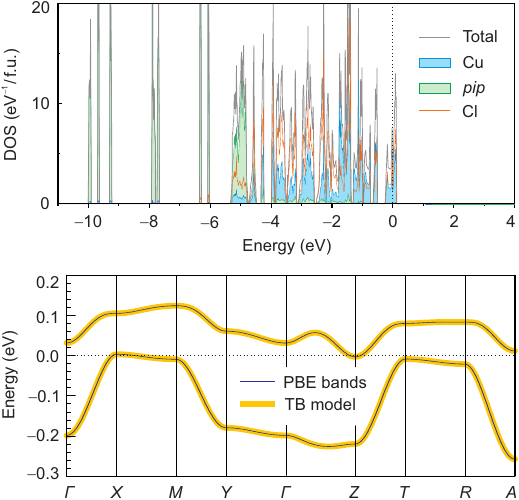}
\caption{\label{fig:bands}
Top: PBE density of states for PHCC, \textit{pip} stands for the contribution of the atoms forming the piperazinium ion. The Fermi level is at zero energy. Bottom: Dispersion of the Cu $d_{x^2-y^2}$ bands calculated in PBE and its fit with the tight-binding (TB) model. The $k$-path is defined as follows: $\Gamma(0,0,0)$, $X(\frac12,0,0)$, $M(\frac12,\frac12,0)$, $Y(0,\frac12,0)$, $Z(0,0,\frac12)$, $T(\frac12,0,\frac12)$, $R(\frac12,\frac12,\frac12)$, $A(0,\frac12,\frac12)$.
}
\end{figure}

The triclinic crystal structure of PHCC is visualized in Fig.~\ref{fig:structure}a. It features Cu$^{2+}$ ions surrounded by five Cl atoms. One out of five Cu--Cl distances (2.606\,\r A) is much longer than the other four ($2.27-2.32$\,\r A), indicating the formation of the plaquette unit typical for the Cu$^{2+}$ ions with the magnetic $d_{x^2-y^2}$ orbital. PBE band structure calculations (Fig.~\ref{fig:bands}, top) confirm this assessment. Apart from the narrow bands arising from the molecular states of the piperazinium ion (below $-5$\,eV), one finds the strongly hybridized Cu $3d$ and Cl $3p$ states above $-4$\,eV. The upper two bands of this Cu--Cl complex cross the Fermi level and should be responsible for the magnetism of PHCC. Those bands are dominated by the Cu $d_{x^2-y^2}$ orbital. 

Geometrically, PHCC features the 1D structural network of Cu$^{2+}$ ions, with the chains running along the $c$ direction. These chains formed by the two shortest Cu--Cu distances of 3.422\,\r A ($J_c$) and 3.446\,\r A ($J_1$), respectively, led the authors of Ref.~\cite{daoud1986} to use the model of bond-alternating spin chain to analyze magnetic susceptibility data. Later neutron-scattering work showed the importance of longer-range interactions, and two possible models with either 6 or 8 nonequivalent couplings were proposed~\cite{stone2001}. 

\begin{table}
\caption{\label{tab:exchange}
Cu--Cu distances in PHCC, hopping parameters of the tight-binding model $t_i$, and exchange couplings $J_i$ obtained from the DFT+$U$ mapping analysis with three representative values of the Coulomb repulsion parameter $U_d$ and the Hund coupling $J_d=1$\,eV. All further couplings are below 0.5\,K.
}
\begin{ruledtabular}
\begin{tabular}{ccr@{\hspace{0.5em}}c@{\hspace{-0.5em}}rrr}
 & $d_{\rm Cu-Cu}$ (\r A) & $t_i$ (meV) & & \multicolumn{3}{c}{$J_i$ (K)} \\\medskip
    & & & $U_d$ (eV): &  8.5  &  9.5 & 10.5 \\ \hline
$J_c$  & 3.422 & 2  & & $-10$ & $-7$ & $-6$ \\
$J_1$  & 3.446 & 87 & &  58   &  44  & 32   \\
$J_1'$ & 6.705 & 33 & &  17   &  14  & 10   \\
$J_2$  & 7.971 & $-35$ & & 20 &  16  & 12   \\
$J_{\perp}$ & 8.274 & 0 & & 4 &   3  &  2   \\
\end{tabular}
\end{ruledtabular}
\end{table}

Here, we assess the magnetic couplings in PHCC using two complementary methods. On the one hand, we obtain the $J_i$ values from the mapping analysis with $U_d=8.5-10.5$\,eV and $J_d=1$\,eV as the typical DFT+$U$ parametrization for Cu$^{2+}$~\cite{liang2025,panther2023,mazurenko2014}. On the other hand, we use hoppings $t_i$ of the tight-binding model for the Cu $d_{x^2-y^2}$ bands (Fig.~\ref{fig:bands}, bottom) to assess the antiferromagnetic (AFM) part of the exchange, $J_i^{\rm AFM}\sim t_i^2$. The results from both methods are listed in Table~\ref{tab:exchange}.

Both short-range couplings, $J_1$ and $J_c$, correspond to the Cu--Cu dimers bridged by two Cl atoms. Despite the very similar Cu--Cu distances, these couplings largely differ in magnitude and even have different sign: while $J_1$ is AFM, $J_c$ is weaker and ferromagnetic (FM). This is because $J_1$ involves two Cl atoms of the CuCl$_4$ plaquette, whereas $J_c$ involves the apical Cl atoms, which are not part of the plaquette. The Cu--Cl--Cu bond angles of $96.0^{\circ}$ for $J_1$ are in the typical range for spin-dimer magnets, such as TlCuCl$_3$ ($95.6^{\circ}$)~\cite{cavadini2001,oosawa2002}, and support an AFM coupling. 

The $J_1$ spin dimers are connected by the long-range couplings $J_1'$ and $J_2$. Both couplings are sizable despite the extended Cu--Cu contacts of $7-8$\,\r A, thanks to the Cu--Cl--Cl--Cu superexchange mechanism assisted by the spatially extended Cl $3p$ orbitals (Fig.~\ref{fig:structure}c). Similar couplings have been reported in (CuCl)LaNb$_2$O$_7$ and related compounds where the long-range superexchange via two Cl atoms exceeds short-range couplings via a single Cl atom, leading to the formation of spin dimers with the unusually long Cu--Cu intradimer distance of more than 8\,\r A~\cite{tassel2010,tsirlin2010,tsirlin2012}. In PHCC, $J_2$ arises from the efficient orbital overlap along the almost co-aligned Cu--Cl bonds with the Cu--Cl--Cl angle of $162.0^{\circ}$. The coupling $J_1'$ involves two pathways of this type, but they strongly deviate from linearity and together result in a weaker coupling compared to $J_2$. 

The two leading couplings $J_1$ and $J_2$ form two-leg spin ladders frustrated by the diagonal coupling $J_1'$. Interestingly, this coupling runs along one of the ladder diagonals only. The coupling along the second diagonal is negligible because of the much longer Cl--Cl distance of 5.91\,\r A (vs. 3.67\,\r A in the case of $J_1'$) that prevents any orbital overlap along the Cu--Cl--Cl--Cu pathway. 
The resulting 1D coupling geometry is known as the `zigzag ladder' model but it can also be viewed as a frustrated dimerized (zigzag) chain, with the alternating nearest-neighbor couplings $J_1$ and $J_1'$ and a $J_2$ coupling between second neighbors. The physics of this model in various regimes of the parameter space is well understood thanks to many previous works~\cite{Harris1973,White1996,Brehmer1996,Oitmaa1996,Uhrig1996,Uhrig1997,Pati1997,Brehmer1998,Totsuka1998,Mila1998,Bouzerar1998,Barnes1999,Singh1999,Mori1999,Kotov1999a,Kotov1999b,Muller2000}, and is reviewed in Ref.~\cite{Mikeska2004}. 

The zigzag chains are further linked in the $ac$ plane by the FM and nonfrustrated coupling $J_c$, whereas the weak AFM coupling $J_{\perp}$ connects the resulting layers along the $b$-direction. Together they lead to a field-induced magnetic order with the propagation vector of $(\frac12,0,\bar{\frac12})$, in agreement with the experiment~\cite{stone2007}. This magnetic order does not appear in zero field because of the gapped nature of the {\jjj} zigzag chain, but it is stabilized in the applied fields above 7.5\,T once the gap is closed.

Our microscopic magnetic model of PHCC is somewhat similar to the 6-parameter model proposed by Stone \textit{et al.}~\cite{stone2001} and later elaborated upon by Perren \textit{et al.}~\cite{perren2015}, who also found the leading AFM couplings $J_1$ and $J_2$ augmented by the FM $J_c$. However, that model features only a minute $J_1'$, thus lacking the frustration of the {\jjj} chain. Moreover, most of the previous literature has relied on the 8-parameter model of Ref.~\cite{stone2001} that includes one of the strong AFM couplings between Cu ions which are 10.3\,\r A apart. We do not find such a coupling in our DFT analysis. Moreover, microscopic arguments suggest that no orbital overlap would be possible for such a long pathway where the Cl--Cl distance must exceed 5\,\r A. 


\section{Fits to magnetic susceptibility and magnetization process data}
\begin{figure}[tb]
\includegraphics[width=8.6cm]{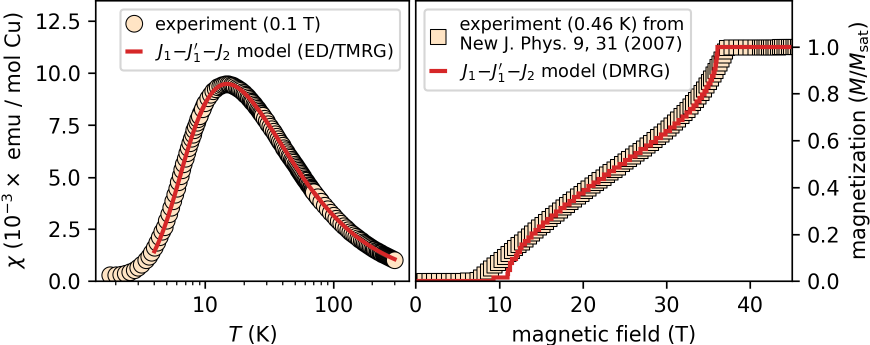}
\caption{\label{fig:fit}
Left: Fit of the experimental $\chi(T)$ using Eq.~\eqref{eq:fit} with the ED/TMRG-simulated magnetic susceptibility of the {\jjj} model with $J_1\!=\!23.1$\,K, $J_1'\!=\!7.0$\,K, $J_2\!=\!13.9$\,K, $g\!=\!2.023$, and $\chi_0\!=\!-1.6\times10^{-4}$ emu\,(mol\,Cu)$^{-1}$.  Right: Fit of the experimental magnetization $M(H\!\parallel\!b)$ from Ref.~\cite{stone2007} using Eq.~(\ref{eq:mh}) with the DMRG-simulated magnetization, $g_b\!=\!2.1$, and the exchange parameters from the $\chi(T)$ fit.}
\end{figure}

Based on our DFT estimates, we initially set $J_1'$ and $J_2$ to 30\% of $J_1$, and calculate the reduced magnetic susceptibility $\chi^*$ defined by 
\be\label{eq:fit}
\chi(T) = \frac{N_{\text{A}}g^2\mu_{\text{B}}^2}{J_1k_{\text{B}}}\chi^*(t^\ast) + \chi_0\,,
\ee
using TMRG and ED. Here $t^\ast\!=\!T/J_1$, $N_{\text{A}}$, $\mu_{\text{B}}$, and $k_{\text{B}}$ are, respectively, the Avogadro number, the Bohr magneton, and the Boltzmann constant. We compare with the experimental $\chi(T)$ measured on a powder sample at 0.1\,T~\cite{chit}, and treat $J_1$ (in K), $g$ and $\chi_0$ as fitting parameters. Despite a rough overall agreement with the experimental data (not shown), least-squares fits reveal visible deviations at high temperatures, while the fitted $g$ = 1.877 and $\chi_0 = 2.2\times10^{-4}$ emu\,(mol\,Cu)$^{-1}$  disagree with the respective estimates from a Curie-Weiss fit above 150\,K: $g$ = 2.036 and $\chi_0 = -2.0\times10^{-4}$ emu\,(mol\,Cu)$^{-1}$. To improve the agreement, we vary $J_1'/J_1$ and $J_2/J_1$ ratios in a reasonable range. In this way, we find that the solution $J_1=23.1$\,K, $J_1'=7.0$\,K, and $J_2=13.9$\,K ($J_1\!:\!J_1'\!:\!J_2\!=\!1\!:\!0.3\!:\!0.6$) is in excellent agreement with the experimental magnetic susceptibility down to 4\,K (Fig.~\ref{fig:fit}, left). Moreover, the theoretical Weiss temperature $\theta_{\text{W}}\!=\!-\frac14 (J_1+J_1'+2J_2)\!=\!-14.5$\,K 
matches well with the experimental $\theta_{\text{W}}=-15.4$\,K, while the fitted $g=2.023$ and $\chi_0=-1.6\times10^{-4}$ emu\,(mol\,Cu)$^{-1}$ are very close to the corresponding estimates from the Curie-Weiss fit.

To get additional confidence in the estimated exchange parameters, we turn to the magnetization curve $M(H)$. The ground-state magnetization $m^*(h^*)$ [see Eq.~(\ref{eq:mh}) below] of the {\jjj} model can be efficiently simulated using DMRG. For an experimental reference, we take the 0.46\,K isotherm measured on single crystals with a magnetic field along the crystallographic $b$ axis (data shown Fig.~2 of Ref.~\cite{stone2007}). Due to the anisotropy of the $g$-factor, we cannot use the $g$ value from our $\chi(T)$ fits which are based on powder measurements. Instead, we treat $g_b$ as a single adjustable parameter in the following expression:
\be\label{eq:mh}
M/M_{\text{sat}}(H) = 2m^*\left(\frac{J_1k_{\text{B}}}{g_b\mu_{\text{B}}}h^*\right)\,.
\ee
The result for $g_b=2.1$ is plotted in Fig.~\ref{fig:fit} (right). The overall agreement is very good, except for the field range around the gap closing. This discrepancy is related to the interchain exchange $J_c$ (see Table~\ref{tab:exchange}), which leads to a reduction of the spin gap. Naturally, our model does not account for this effect due to its 1D nature. Notably, $J_c$ is ferromagnetic and hence it should not affect the saturation field -- indeed, experimental and simulated curves agree very well at the verge of saturation.

Table~\ref{tab:Hcs} shows a comparison between the critical fields $H_{\text{c}1}$ and $H_{\text{c}2}$ reported in Ref.~\cite{stone2007} and the theoretical values obtained from three different approaches (all for an isolated chain) based on the parameter set obtained from the susceptibility and the magnetization process fits: i) first-order perturbation theory around the isolated-rungs limit~(see Refs.~\cite{Uhrig1997,Mila1998} and Eq.~(\ref{eq:Hc1}) in App.~\ref{app:PT}), ii) third-order perturbation theory~\cite{Uhrig1997}, and iii) DMRG calculations. 
As mentioned above, the agreement for $H_{\text{c}1}$ is satisfactory and the small deviation by a few T can be attributed to the interchain triplet hopping due to $J_\perp$ and $J_c$. 
Turning to the saturation field $H_{\text{c}2}$, we note that, for an isolated chain, the first-order expression for the saturation field, 
\be
g \mu_B H_{\text{c}2} = J_1+2J_2\,,
\ee
is exact to all orders~\cite{Mila1998}, and, as mentioned above, remains valid in the presence of the ferromagnetic interchain coupling $J_c$. The slight disagreement in the value of $H_{\text{c}2}$ can be attributed to the much weaker interchain coupling $J_\perp$. 

Altogether, the {\jjj} model describes the magnetic susceptibility and field-dependent magnetization data very accurately, Any deviations due to the weaker interchain couplings are small and can be incorporated at a mean-field level.  
The exchange couplings extracted from the fits place PHCC in the regime of strong dimerization ($J_1'\!\ll\!J_1$) and relatively close to the regular ladder model ($J_1\!=\!J_2$, $J_1'\!=\!0$), and, in particular, in the XY-universality class of the model~\cite{Mila1998}.

\begin{table}[t]
\caption{Comparison between the experimental values of the critical fields $H_{\text{c}1}$ and $H_{\text{c}2}$ reported in Ref.~\cite{stone2007} for ${\bf H}\parallel {\bf b}$ (second column), and the theoretical values obtained from 1st-order perturbation theory (third column), 3rd-order perturbation theory (fourth column)~\cite{Uhrig1997}, and DMRG calculations (fifth column), for the parameter set obtained from the susceptibility and the magnetization process fits ($J_1=23.1$\,K, $J_1'=7.0$\,K, $J_2=13.9$\,K, $g_b$ = 2.1).}\label{tab:Hcs}
\ra{1.35}\setlength{\tabcolsep}{1pt}
\begin{tabular}{@{\extracolsep{\fill}} l c c c c}
\toprule
$\ra{0.5}\begin{array}{c}\text{critical}\\\text{fields}\end{array}$
& 
$\ra{0.5}\begin{array}{c}\text{Experiment}\\\text{\cite{stone2007}}\end{array}$
&
$\ra{0.5}\begin{array}{c}\text{1st-order PT}\\\text{\cite{Uhrig1997,Mila1998}, App.~\ref{app:PT}}\end{array}$
&
$\ra{0.5}\begin{array}{c}\text{3rd-order}\\\text{\cite{Uhrig1997}}\end{array}$
& DMRG\\
\midrule
$H_{\text{c}1}$ [T] & 7.5 & 9& 11.8 &  9.3\\
$H_{\text{c}2}$ [T] & 37 & 36.1& 36.1& 36.1\\
\bottomrule
\end{tabular}
\end{table}

\section{Excitation spectrum}
Having established the minimal microscopic model and a reliable set of exchange parameters, we now turn to the structure of the excitation spectrum, in order to compare with published INS data and make further predictions for PHCC. In particular, we aim to check whether the above minimal model explains the experimental observation of triplets merging into the two-particle continuum~\cite{stone2006}, but also explore whether PHCC manifests any bound and anti-bound states of pairs of triplets, which are known to exist in certain parameter regions of spin-1/2 ladders and similar models~\cite{Uhrig1996,Pati1997,Bouzerar1998,Barnes1999,Kotov1999a,Kotov1999b,Mori1999,Zheng2001,Nayak2020}. 

To address these questions we first perform numerical exact diagonalizations on $24$-site clusters with periodic boundary conditions. Since the Heisenberg Hamiltonian commutes with $S^z$, the diagonalization can be done separately in each $S^z$ sector. In Fig.~\ref{fig:spectrum}, we plot the 30 lowest-lying states with $S^z\!=\!0-3$ as a function of $k$. 

\subsection{Single-triplets vs two-particle continuum}
\begin{figure}[!t]
\includegraphics[width=8.6cm]{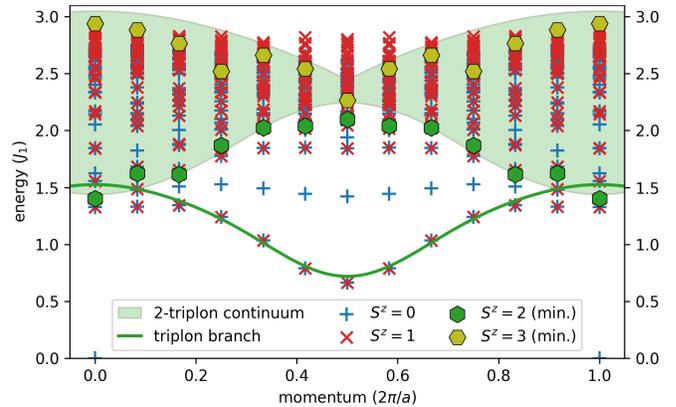}
\caption{Excitation spectrum of the {\jjj} model with $J_1:J_1':J_2=1:0.3:0.6$ calculated for the $24$-site chain with periodic boundary conditions. We show 30(1) lowest-lying excitations for $S^z=0,1$ ($S^z=2,3$) sectors. The solid curve is the excitation energy $\omega_{1t}^{(3)}(k)$ of the single-triplet band according to third-order perturbation theory~\cite{Uhrig1997}. The shaded (green) region is the corresponding (non-interacting) two-particle continuum delimited by the boundaries  $\omega^{(3)}_{\text{2t,min}}(k) = \min_{q} \big\{ \omega^{(3)}_{1t}(q) + \omega^{(3)}_{1t}(k-q)\}$ and $\omega^{(3)}_{\text{2t,max}}(k) = \max_{q} \big\{ \omega^{(3)}_{1t}(q) + \omega^{(3)}_{1t}(k-q) \}$}.
\label{fig:spectrum}
\end{figure}

The results of Fig.~\ref{fig:spectrum} show a coherent triplet band with a minimum at the boundary of the first Brillouin zone ($k=\pi/a$), followed by a thicket of higher-lying excitations -- indicative of a two-triplon continuum. 
Additionally, the triplet band appears to merge into the continuum in the vicinity of the $\Gamma$-point ($k\!=\!0$).

To obtain a more direct connection with the triplet picture we compare the ED spectra with the theoretical predictions by Uhrig~\cite{Uhrig1997} (based on third-order perturbation theory around the isolated rungs limit) for the single-triplet band $\omega^{(3)}_{1t}$ and the corresponding boundaries, $\omega^{(3)}_{\text{2t,min}}$ and $\omega^{(3)}_{\text{2t,max}}$, of the (non-interacting) two-particle continuum. 
The agreement for the one-particle band is very satisfactory, confirming the assertion~\cite{Uhrig1997} that perturbation theory provides accurate results even in the regime of large alternation between $J_1$ and $J_1'$, which is the case of PHCC. 
For the two-particle continuum, the third-order expansion captures the lowest energy region well (boundary close to the $\Gamma$ point) but deviations from the ED data become evident at higher energies, including the lower boundary of the continuum around $k\!=\!\pi$ and the upper boundary of the continuum (where we also observe $S^z\!=\!3$ states merging into the two-particle continuum). 

Taken together, the spectral results are consistent with the experimental data reported by Stone {\it et al.}~\cite{stone2007}, lending further strong support for the validity of the minimal {\jjj} model.

\subsection{Two-particle bound states}
Another notable feature of the excitation spectrum shown in Fig.~\ref{fig:spectrum} is the existence of a weakly-dispersive singlet band (for $\pi/6\lesssim k \lesssim 11\pi/6$, around $\sim1.5J_1$) and an almost flat triplet band (for $\pi/2\lesssim k \lesssim 3\pi/2$, around $\sim1.8J_1$). These bands appear below the two-particle continuum and resemble the bound states discussed in spin-1/2 ladders and similar models~\cite{Uhrig1996,Pati1997,Bouzerar1998,Barnes1999,Kotov1999a,Kotov1999b,Mori1999,Zheng2001,Nayak2020}. 

\begin{figure}[!t]
\centering
~~~~
\includegraphics[width=0.96\columnwidth]{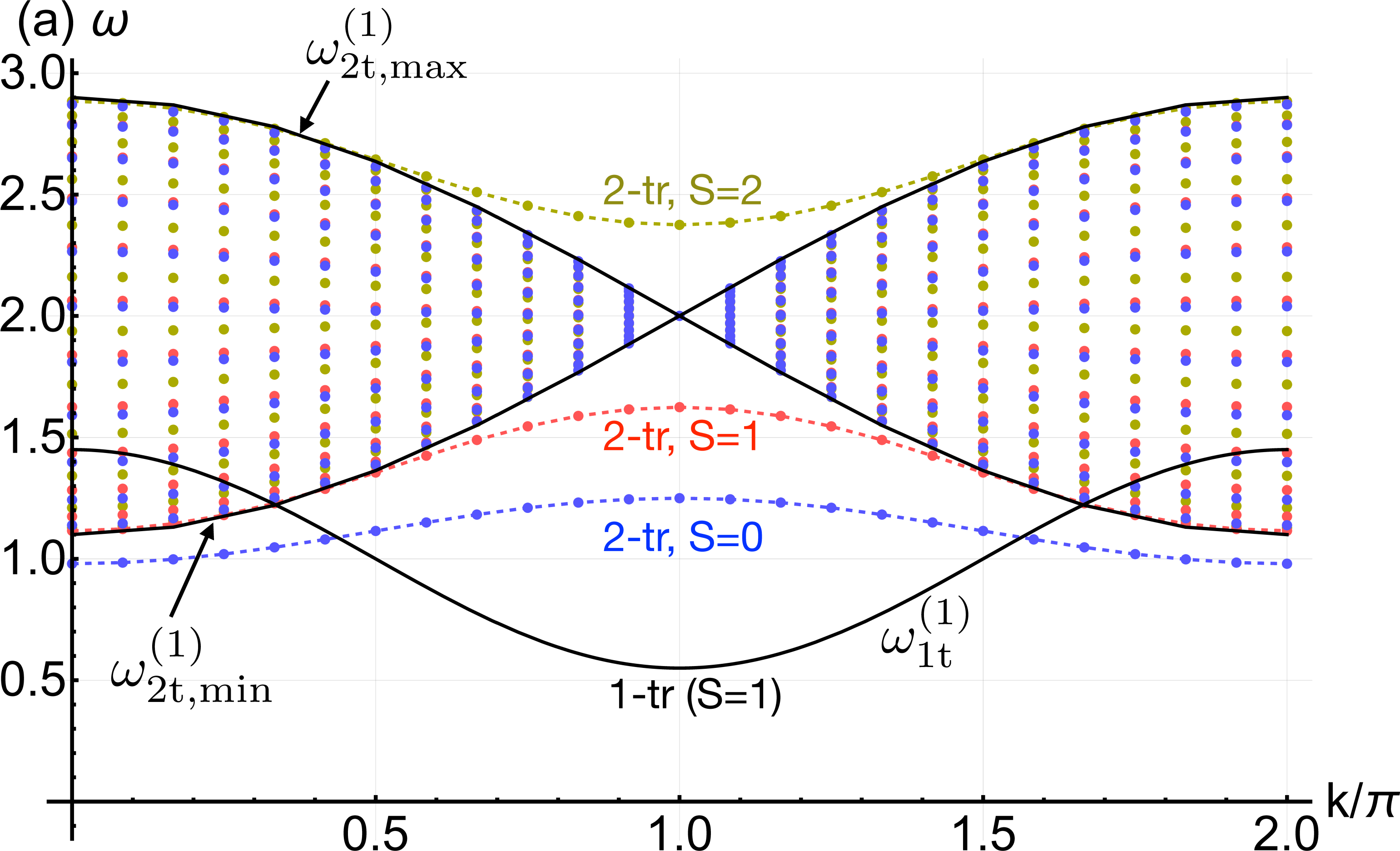}
\includegraphics[width=0.96\columnwidth]{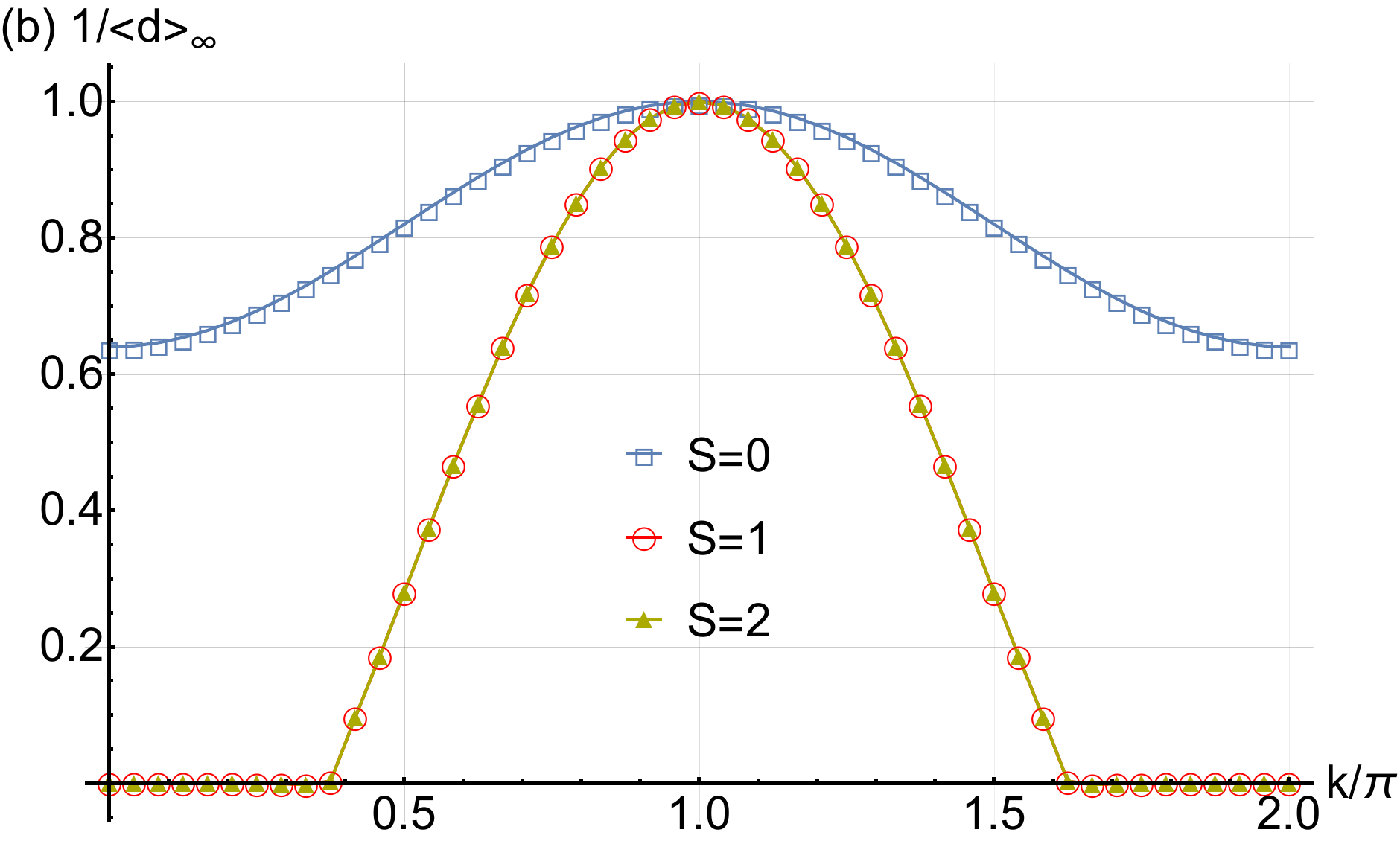}
\caption{(a) Excitation energies of two-particle states of the {\jjj} model on a system with 24 rungs and parameters $J_1=1$, $J_1'=0.3$ and $J_2=0.6$, as obtained from a numerical diagonalization of the matrix representation of $\mc{H}$ inside the two-particle sector [see Eq.~(\ref{eq:Vprojected})]. 
The two-particle states with $S=0$, $1$ and $2$ are shown by blue, red and dark yellow symbols, respectively.
The solid black lines show the first-order single-triplon energy $\omega_{1t}^{(1)}(k)$ of Eq.~(\ref{eq:w1ofk}) and the corresponding predictions for the boundaries $\omega^{(1)}_{2\text{t,min}}$ and $\omega^{(1)}_{2\text{t,max}}$, of the non-interacting two-particle continuum. 
(b) Extrapolated values of the inverse of the coherence length $\langle\hat{d}\rangle$ of the lowest $S\!=\!0$, the lowest $S\!=\!1$ and highest $S\!=\!2$ bands, as obtained from diagonalizations of the Hamiltonian matrix inside the two-particle sector (see App.~\ref{app:PT}), and using finite-size extrapolations of $1/\langle\hat{d}\rangle$ with system sizes with $N_r\!=\!48, 56, \cdots,160$ rungs.
}
\label{fig:PT}
\end{figure}

These features can be explained qualitatively already at the level of first-order perturbation theory. 
The analysis is presented in App.~\ref{app:PT} and the main results are presented in Fig.~\ref{fig:PT}. In particular, panel (a) shows the energies of the two-particle states resulting from a numerical diagonalization of the  Hamiltonian matrix inside the two-particle sector for a system of $N_r\!=\!24$ rungs and parameters $J_1\!=\!1$, $J_1'\!=\!0.3$ and $J_2\!=\!0.6$. The spectrum includes the energies of all $S\!=\!0$, $S\!=\!1$ and $S\!=\!2$ two-particle states, and we have also included the first-order results for the single-particle energy $\omega_{1t}^{(1)}(k)$ [see Eq.~(\ref{eq:w1ofk})] as well as the corresponding boundaries of the (non-interacting) two-particle scattering continuum, $\omega^{(1t)}_{2\text{t,min}}$ and $\omega^{(1t)}_{2\text{t,max}}$, as obtained from $\omega_{1t}^{(1)}(k)$.
In Fig.~\ref{fig:PT}\,(b) we present the so-called `coherence length'~\cite{Zheng2001} [the average distance between the two involved triplets, see Eq.~(\ref{eq:CL})] for the lowest singlet, the lowest triplet and the highest quintuplet of the two-particle energies shown in Fig.~\ref{fig:PT}\,(a). 
The key results can be summarized as follows: 

i) Nearest-neighbour (NN) triplets forming a singlet attract each other with an amplitude $-(J_2+J_1'/2)$. As seen in the spectra of Fig.~\ref{fig:PT}\,(a), this gives rise to a coherent band of singlets below the bottom edge of the two-particle continuum. The coherence length shown in Fig.~\ref{fig:PT}\,(b) remains finite and short for all momenta, confirming that these are bound states.  

ii) Nearest-neighbour (NN) triplets forming a triplet attract each other with an amplitude $-(J_2+J_1'/2)/2$. The lowest triplet shown in Fig.~\ref{fig:PT}\,(a) is below the bottom edge of the continuum only in a finite momentum region around $k\!=\!\pi$. The coherence length shown in Fig.~\ref{fig:PT}\,(b) is finite inside this region and infinitely large outside, showing that the former correspond to bound states and the latter to scattering states. 

iii) Nearest-neighbour (NN) triplets forming a quintent repel each other with an amplitude $+(J_2+J_1'/2)/2$, opposite to the $S\!=\!1$ binding amplitude. As a result, the spectrum of quintuplets is the mirror image of the spectrum of triplets around the zeroth-order energy $2J_1$, see Fig.~\ref{fig:PT}\,(a). The highest energy quintuplet is the mirror of the lowest triplet, and its part that lies above the upper edge of the continuum corresponds to anti-bound states, as also shown by the coherence length of Fig.~\ref{fig:PT}\,(b) (which is identical to that of the lowest two-particle $S\!=\!1$ band).

iv) The scattering continuum forms a bowtie-like structure with all states staying degenerate at $k\!=\!\pi$ with energy $2J_1$.   

The results (i-iv) are valid in first-order in $J_1'$ and $J_2$, and modifications are to be expected at higher orders. A comparison with the ED data of Fig.~\ref{fig:spectrum} shows, in particular, that the part of the singlet band around the $\Gamma$ point is pushed inside the two-particle continuum. Also, the curvature of the singlet band around $k\!=\!\pi$ has opposite sign in the ED data compared to the first-order results. And, finally, the degeneracy at $k\!=\!\pi$ is lifted. This can also be seen from the two-particle (non-interacting) continuum predicted by third-order perturbation theory~\cite{Uhrig1997} (shaded region in Fig.~\ref{fig:spectrum}).

\section{Summary and conclusions}
We propose an accurate microscopic magnetic model for the quantum magnet PHCC (C$_4$H$_{12}$N$_2$)Cu$_2$Cl$_6$~\cite{daoud1986,battaglia1988}, one of the early experimental realisations of a disordered gapped antiferromagnet manifesting a model case scenario of quasiparticle breakdown~\cite{stone2006}. 
Our results show that PHCC is a rare example of the zigzag spin ladder, which can also be viewed as a frustrated dimerized (zigzag) chain (with alternating nearest-neighbor couplings $J_1$ and $J_1'$ and a frustrating second-neighbour coupling $J_2$), in the regime of strong dimerization ($J_1'/J_1\!=\!0.3$ and $J_2/J_1\!=\!0.6$) and in the XY universality class of the model~\cite{Mila1998}.    

The model provides a quantitative description of the magnetic susceptibility and magnetization process. It also accounts for the nature of the ground state, the spin gap, the dispersion of the single-triplet band, and the merging of this band into the two-particle scattering continuum close to the Brillouin zone center, in agreement with experiment~\cite{stone2006}. 

In addition, our numerical results reveal that PHCC should also manifest two-particle bound states of singlets and triplets below the bottom edge of the two-particle continuum, as well as quintuplet anti-bound states above the upper edge of the continuum. The origin of these bound (anti-bound) states can be understood at the level of first-order perturbation theory and originates in the attraction (repulsion) of a pair of triplets sitting on neighbouring rungs, proportional to $J_2\!+\!J_1'/2$.  

The presence of bound states in the calculated energy spectra calls for further studies of PHCC. Most of the inelastic neutron scattering studies are restricted to energies below 3\,meV, whereas we predict the $S=1$ bound triplets around 3.5\,meV. Indeed, some of the energy scans in Ref.~\cite{stone2001} show a weak intensity at $k=\pi$ around this energy, but dedicated $k$-dependent experiments would be necessary to confirm the absence of the dispersion and the bound nature of the corresponding excited state. 

Despite the relative simplicity of the dimerized zigzag chain model, its experimental realizations are scarce. KCuCl$_3$ lies very close to the dimer limit~\cite{cavadini2000}, whereas the spin-Peierls phase of CuGeO$_3$ showcases the regime of weak dimerization with spin and lattice degrees of freedom strongly intertwined~\cite{braden1999,Spitz2025}. This renders PHCC a unique dimerized zigzag chain material with only a moderate dimerization, and motivates its further dedicated investigation.

\acknowledgments
This work is dedicated to the loving memory of Johannes Richter, our long-term collaborator and mentor, to whom we are grateful for many years of fruitful scientific exchange.
\\

We would like to thank Dan H\"uvonen for bringing PHCC to our attention and providing experimental magnetization data for this material. OJ acknowledges financial support from the German Forschungsgemeinschaft (DFG, German Research Foundation) through SFB 1143 (Project ID 247310070) and thanks Ulrike Nitzsche for technical assistance. AT gratefully acknowledges the granted computing time on the high-performance computer at the NHR Center of TU Dresden. This center is jointly supported by the Federal Ministry of Research, Technology and Space of Germany and the state governments participating in the NHR (www.nhr-verein.de/unsere-partner).
IR acknowledges the support by the Engineering and Physical Sciences Research Council, Grant No. EP/V038281/1.

\appendix

\section{Theoretical analysis of the one- and two-particle excitations of PHCC}\label{app:PT}
Several aspects of the structure of the single-triplon band, the two-triplet continuum, and the two-triplet bound and anti-bound states of the dimerized zigzag chain model have been explored in previous works in various parameter regimes~\cite{Harris1973,Oitmaa1996,Uhrig1996,Uhrig1997,Pati1997,Brehmer1998,Bouzerar1998,Barnes1999,Singh1999,Kotov1999a,Kotov1999b,Mori1999,Muller2000,Zheng2001}. 
Here we wish to outline the key aspects for the particular parameter regime of PHCC by combining perturbation theory and exact diagonalizations on finite-size lattices.

\begin{figure}[!t]
\includegraphics[width=0.95\columnwidth]{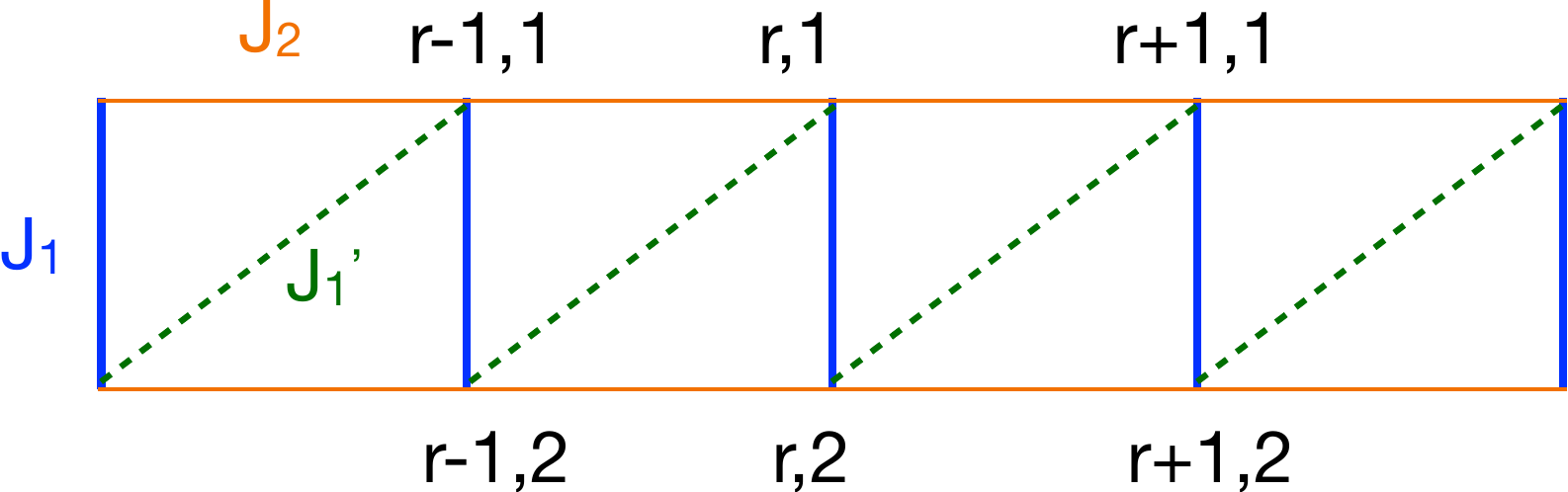}
\caption{The {\jjj} 2-leg ladder model.}\label{fig:model}\end{figure}

\subsection{The Hamiltonian}
We consider an isolated zigzag ladder with $N_r$ rungs and periodic boundary conditions. The spin-1/2 {\jjj} model is described by the Hamiltonian (see Fig.~\ref{fig:model})
\bea
\mc{H}\!&\!=\!&\!\sum_{r=1}^{N_r} \{ J_1 {\bf S}_{r,1}\cdot {\bf S}_{r,2} + J_2 \left( {\bf S}_{r,1}\cdot {\bf S}_{r+1,1} 
+{\bf S}_{r,2}\cdot {\bf S}_{r+1,2} \right) \nonumber\\
&&~~+J_1' {\bf S}_{r+1,1}\cdot {\bf S}_{r,2} \}\,,
\eea
where $r$ labels the rungs and we have set the rungs spacing $a=1$. In the following, we define 
\be
\alpha_{\pm} \equiv J_2\pm J_1/2\,.
\ee 
We wish to expand around the isolated-rung limit $J_1'=J_2=0$. To that end, we denote by 
\be
\mc{H}_0 = \sum_{r}\! J_1 {\bf S}_{r,1}\cdot {\bf S}_{r,2}\,, \\
\ee
the unperturbed Hamiltonian and by $\mc{V}=\mc{H}-\mc{H}_0$ the perturbation. 

\subsection{Ground state}
\begin{figure}[!t]
\includegraphics[width=0.99\columnwidth]{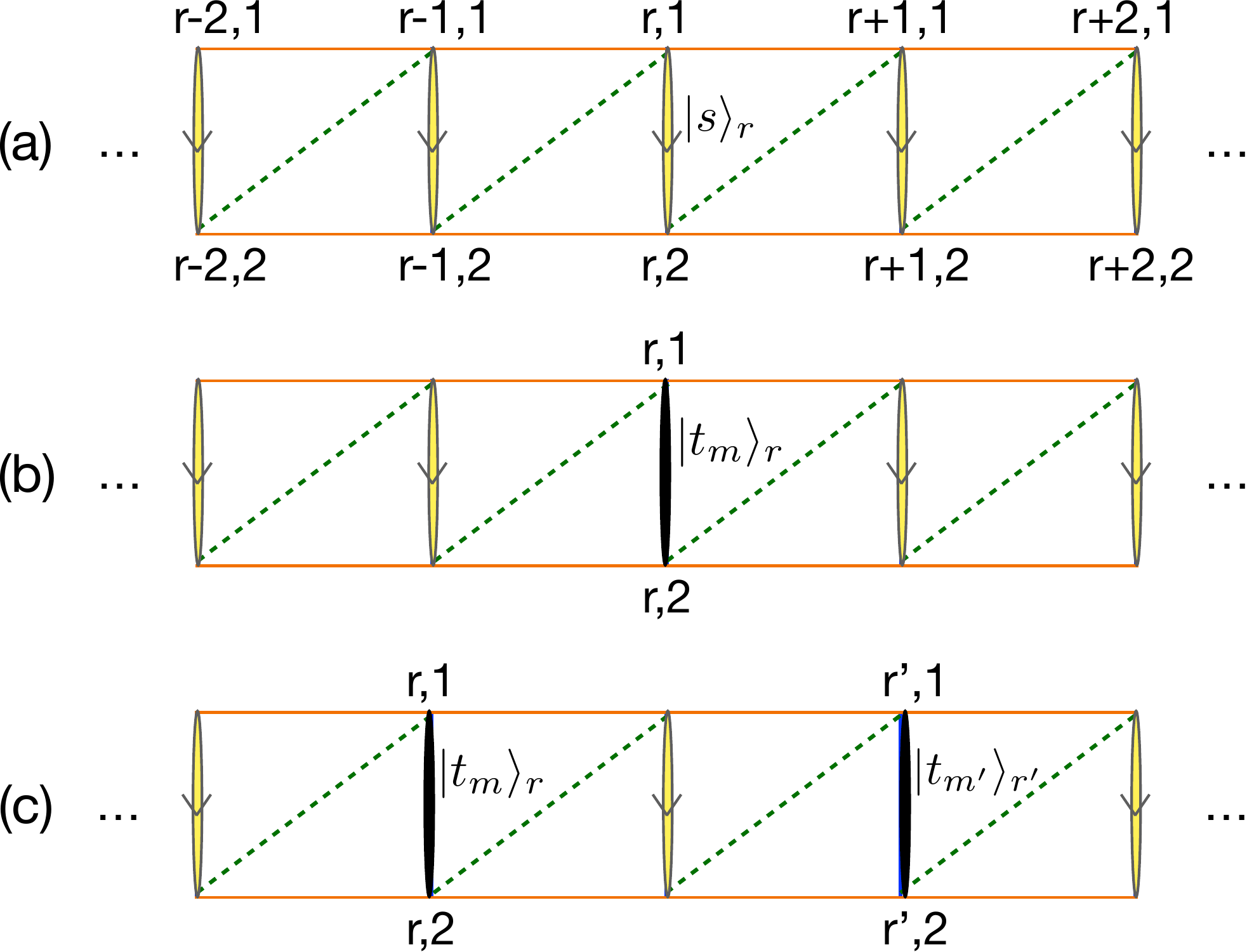} 
\caption{Ground state and elementary single- and two-particle excitations of the isolated-rung model. Rung singlets are shown by yellow ovals, and rung-triplets are shown by black ovals.}
\label{fig:ElemExcitationsH0}\end{figure}

The ground state of $\mc{H}_0$ is the product of singlets on the rungs (see Fig.~\ref{fig:ElemExcitationsH0}\,a)
\be
|\psi_0^{(0)}\rangle = \prod_{r}^\otimes |s\rangle_{r}
=\prod_{r}^\otimes \left(\frac{|\uparrow\downarrow\rangle-|\downarrow\uparrow\rangle}{\sqrt{2}}\right)_{r,1;r,2}\,,
\ee
with energy $E_0^{(0)} = -\frac{3J_1}{4} N_r$. 
To first-order in $J_1'$ and $J_2$, the ground state energy does not change (i.e., $E_0^{(1)} =E_0^{(0)} $), since $\mc{V} |\psi_0^{(0)}\rangle$ is orthogonal to $|\psi_0^{(0)}\rangle$.

\subsection{Single-triplet sector}
The elementary excitations in the isolated-rung limit arise by promoting one of the rung singlets into a triplet (see Fig.~\ref{fig:ElemExcitationsH0}\,b):
\be
|\psi_{m,r}^{(0)}\rangle = |t_m\rangle_r \otimes \prod_{r' \neq r}^\otimes |s\rangle_{r'}\,,
\ee
where
\be
\ra{1.35}
\begin{array}{l}
|t_1\rangle_r \equiv |\uparrow\uparrow_{r,2}\rangle_{r,1; r,2}\,,\\
|t_0\rangle_r \equiv \frac{1}{\sqrt{2}}\left(|\uparrow\downarrow\rangle+|\downarrow\uparrow\rangle\right)_{r,1; r,2}\,,\\
|t_{-1}\rangle_r \equiv |\downarrow\downarrow\rangle_{r,1; r,2}\,.
\end{array}
\ee
These states form a $(3N_r)$-fold degenerate manifold with unperturbed energy $E_{1t}^{(0)}=E_0^{(0)}+J_1$.

To find the first-order correction to the energy of these triplets we must diagonalize the matrix representation of $\mc{H}$ inside the manifold of these states. We have the general relations:
\begin{widetext}
\be
\ra{1.35}
\begin{array}{l}
{\bf S}_{r,2}\cdot {\bf S}_{r',1} ~ |t_1\rangle_r \otimes |s\rangle_{r'} =  
\frac{1}{4} \big[-|s\rangle_r \otimes |t_1\rangle_{r'}+ |t_1\rangle_r \otimes |t_0\rangle_{r'}-|t_0\rangle_r \otimes |t_1\rangle_{r'} \rangle\big]\,,
\\
{\bf S}_{r,2}\cdot {\bf S}_{r',1} ~ |t_{-1}\rangle_r \otimes |s\rangle_{r'} =  
\frac{1}{4} \big[-|s\rangle_r \otimes |t_{-1}\rangle_{r'} - |t_{-1}\rangle_r \otimes |t_0\rangle_{r'}+|t_0\rangle_r \otimes |t_{-1}\rangle_{r'} \rangle\big]\,,
\\
{\bf S}_{r,2}\cdot {\bf S}_{r',1} ~ |t_0\rangle_r \otimes |s\rangle_{r'} =  
\frac{1}{4} \big[-|s\rangle_r \otimes |t_0\rangle_{r'} + |t_1\rangle_r \otimes |t_{-1}\rangle_{r'}-|t_{-1}\rangle_r  \otimes |t_1\rangle_{r'} \rangle\big]\,.
\end{array}
\ee
\end{widetext}
Repeating for other types of couplings and disregarding the terms that do not belong to the manifold of single-triplet states we obtain
\be
\ra{1.35}
\begin{array}{l}
{\bf S}_{r,2}\cdot {\bf S}_{r',1} ~ |t_m\rangle_r \otimes |s\rangle_{r'} =  -\frac{1}{4} |s\rangle_r \otimes |t_m\rangle_{r'}\,,
\\
{\bf S}_{r,2}\cdot {\bf S}_{r',1} ~ |s\rangle_r \otimes |t_m\rangle_{r'} =  -\frac{1}{4} |t_m\rangle_r \otimes |s\rangle_{r'}\,,
\\
{\bf S}_{r,\nu}\cdot {\bf S}_{r',\nu} ~ |t_m\rangle_r \otimes |s\rangle_{r'} =  
+\frac{1}{4} |s\rangle_r  \otimes |t_m\rangle_{r'}\,,
\end{array}
\ee 
for all $m=-1,0,1$ and $\nu=1,2$. Hence, we find that  (again disregarding wavefunction components outside the single-triplet manifold)
\be\label{eq:Vpsimr}
\mc{V}~|\psi_{m,r}^{(0)}\rangle  = \frac{\alpha_-}{2} \left(|\psi_{m,r+1}^{(0)}\rangle +|\psi_{m,r-1}^{(0)}\rangle\right)\,.
\ee
Going into momentum space
\be
|\psi_{m,k}^{(0)}\rangle \equiv \sum_{r} e^{i k r} ~|\psi_{m,r}^{(0)}\rangle\,,
\ee
where $k$ is the momentum along the legs of the ladder, we obtain
\be
\mc{H}~|\psi_{m,k}^{(0)}\rangle  = E_{1t}^{(1)}~|\psi_{m,k}^{(0)}\rangle
\ee
where $E_{1t}^{(1)}=E_0^{(1)} + \omega_{1t}^{(1)}(k)$, and  
\be\label{eq:w1ofk}
\omega_{1t}^{(1)}(k) = J_1 + \alpha_- \cos{k}\,,
\ee
which agrees with published results~\cite{Uhrig1997,Mila1998}. 
Note that, for $\alpha_-=0$ (i.e., $J_2=J_1'/2$), the band is completely flat with energy $\omega_{1t}^{(1)}(k) = J_1$. This result is only true to leading order in $J_2$ and $J_1'$.

Uhrig~\cite{Uhrig1997} has obtained analytical expressions up to third-order in $J_1'$ and $J_2$ (which, as discussed in the main text, are in very good agreement with our ED data of Fig.~\ref{fig:spectrum}), and there are published results from higher-order series expansions at various points in parameter space~\cite{Oitmaa1996,Barnes1999,Singh1999}.

The critical field $H_{\text{c}1}$ corresponds to the closing of the gap of the $k=\pi$ mode. To first-order in $J_1'$ and $J_2$,
\be\label{eq:Hc1}
g \mu_B H_{\text{c}1}^{(1)} = \omega_{1t}^{(1)}(\pi)  = J_1 -J_2+J_1'/2 \,, 
\ee
which agrees with published results~\cite{Uhrig1997,Mila1998}.


\subsection{The two-triplet sector} 
Consider the two-triplet states 
\be
|\psi_{r, r'; m,m'}^{(0)}\rangle\!=\! |t_m\rangle_r\otimes |t_{m'}\rangle_{r'}\otimes
\!\!\prod_{r'' \neq r, r'}^\otimes \!|s\rangle_{r''}\,,
\ee
which are formed by promoting the rung-singlets at positions $r$ and $r'$ into triplets (see Fig.~\ref{fig:ElemExcitationsH0}\,c). The zeroth-order energy of these two-triplet states is 
\be
E_{2t}^{(0)} = E_{0}^{(0)}+2J_1\,.
\ee 
The two triplets can combine to give a total singlet, a triplet and a quintuplet. The singlet is given by:
\be 
\!\!|\psi_{r, r'; S=0,M=0}^{(0)}\rangle\!=\!\frac{|\psi_{r,r';1,-1}^{(0)}\rangle \!+\!|\psi_{r,r'; -1,1}^{(0)}\rangle\!-\!|\psi_{r,r'; 0,0}^{(0)}\rangle}{\sqrt{3}}\,.
\ee
The three components of the triplet are given by 
\be
\ra{1.35}
\!\!\begin{array}{l}
|\psi_{r, r'; S=1,M=1}^{(0)}\rangle\!=\!\frac{1}{\sqrt{2}}\big(|\psi_{r,r';1,0}^{(0)}\rangle\!-\!|\psi_{r,r'; 0,1}^{(0)}\rangle \big)\,, 
\\
|\psi_{r, r'; S=1,M=0}^{(0)}\rangle\!=\!\frac{1}{\sqrt{2}}\big(|\psi_{r,r';1,-1}^{(0)}\rangle\!-\!|\psi_{r,r'; -1,1}^{(0)}\rangle \big)\,, 
\\
|\psi_{r, r'; S=1,M=-1}^{(0)}\rangle\!=\!\frac{1}{\sqrt{2}}\big(|\psi_{r,r';0,-1}^{(0)}\rangle\!-\!|\psi_{r,r'; -1,0}^{(0)}\rangle \big)\,, 
\end{array}
\ee
and the five components of the quintuplet are given by
\be
\ra{1.35}
\!\!\!\!\begin{array}{l}
|\psi_{r, r'; S=2,M=2}^{(0)}\rangle\!=\!|\psi_{r,r';1,1}^{(0)}\rangle\,, 
\\
|\psi_{r, r'; S=2,M=1}^{(0)}\rangle\!=\!\frac{1}{\sqrt{2}}\big(|\psi_{r,r';0,1}^{(0)}\rangle\!+\!|\psi_{r,r'; 1,0}^{(0)}\rangle \big)\,, 
\\
|\psi_{r, r'; S=2,M=0}^{(0)}\rangle\!=\!\frac{|\psi_{r,r';1,-1}^{(0)}\rangle+|\psi_{r,r';-1,1}^{(0)}\rangle+2|\psi_{r,r'; 0,0}^{(0)}\rangle }{\sqrt{6}}\,, 
\\
|\psi_{r, r'; S=2,M=-1}^{(0)}\rangle\!=\!\frac{1}{\sqrt{2}}\big(|\psi_{r,r';0,-1}^{(0)}\rangle\!+\!|\psi_{r,r'; -1,0}^{(0)}\rangle \big)\,, 
\\
|\psi_{r, r'; S=2,M=-2}^{(0)}\rangle\!=\!|\psi_{r,r+d;-1,-1}^{(0)}\rangle\,. 
\end{array}
\ee
Let us now obtain the matrix elements of $\mc{V}$ inside the two-triplet sector. 
To that end, we first re-write $r'=r+d$, where $d=1,2,\cdots, d_{\text{max}}$ is the distance between the two triplets. Due to the periodic boundary conditions, $d_{\text{max}}=N_r/2$ if $N_r$ is even and $(N_r-1)/2$ if $N_r$ is odd. Then, we use Eq.~(\ref{eq:Vpsimr}) and disregard wavefunction components outside the two-triplet manifold. Also, we must treat $d=1$ and $d>1$ separately, because the former case includes NN interactions (diagonal matrix-elements), besides hopping terms (off-diagonal terms).
For $d\neq 1$, we find
\bea
&&\!\!\!\mc{V} |\psi_{r, r+d; m,m'}^{(0)}\rangle \!=\!
\frac{\alpha_-}{2} \big[
|\psi_{r, r+d+1; m,m'}^{(0)}\rangle 
\!+\!|\psi_{r, r+d-1; m,m'}^{(0)}\rangle \nonumber\\
&&~~~~~+|\psi_{r+1, r+d; m,m'}^{(0)}\rangle
\!+\!|\psi_{r-1, r+d; m,m'}^{(0)}\rangle
\big],~~
\eea
which, in turn, gives
\bea
&&\!\!\!\mc{V} |\psi_{r, r+d; S,M}^{(0)}\rangle \!=\! 
\frac{\alpha_-}{2} \big[
|\psi_{r, r+d+1; S,M}^{(0)}\rangle 
\!+\!|\psi_{r, r+d-1; S,M}^{(0)}\rangle\nonumber\\
&&~~~~~+|\psi_{r+1, r+d; S,M}^{(0)}\rangle
\!+\!|\psi_{r-1, r+d; S,M}^{(0)}\rangle
\big]\,,
\eea
where we note that for $d=d_{\text{max}}$, $d+1$ is equivalent to $d_{\text{max}}-1$, because of periodic boundary conditions. 
For $d=1$, we find:
\bea
&&\!\!\!\mc{V} |\psi_{r, r+1; S,M}^{(0)}\rangle \!=\! 
\frac{\alpha_-}{2} \big[|\psi_{r, r+2; S,M}^{(0)}\rangle 
+|\psi_{r-1, r+1; S,M}^{(0)}\rangle\big]\nonumber\\
&&~~~~~~~~~~~~~~~~+\beta_S |\psi_{r, r+1; S,M}^{(0)}\rangle \,,
\eea
where
\be
\beta_0=-\alpha_+\,,~~~
\beta_1=-\alpha_+/2\,,~~~
\beta_2=+\alpha_+/2\,.~~~
\ee
Hence, we find that NN triplets forming a singlet attract each other with amplitude $-\alpha_+$, NN triplets forming a triplet attract with amplitude $-\alpha_+/2$, and NN triplets forming a quintuplet repel each other with amplitude $\alpha_+/2$.   
We now go to momentum space and define
\be
|\psi_{k,d; S,M}^{(0)}\rangle \equiv  e^{i k \frac{d}{2}}\frac{1}{\sqrt{N_r}} \sum_{r=1}^{N_r} e^{i k r}|\psi_{r, r+d; S,M}^{(0)}\rangle \,,
\ee
where $k$ is the momentum corresponding to the `center of mass' of the two triplets. We find the following relations:
\be
\ra{1.35}
\!\!\!\!\begin{array}{l}
 d\!\neq\!1\!:
\mc{V} |\psi_{k,d; S,M}^{(0)}\rangle  
\!=\!\alpha_k \big(  |\psi_{k,d+1; S,M}^{(0)}\rangle\!+\! |\psi_{k,d-1; S,M}^{(0)}\rangle\big)\,,
\\
d\!=\!1\!:
\mc{V} |\psi_{k,1; S,M}^{(0)}\rangle  
\!=\! \alpha_k |\psi_{k, 2; S,M}^{(0)}\rangle \!+\!\beta_S |\psi_{k, 1; S,M}^{(0)}\rangle\,.
\end{array}
\ee
where we defined $\alpha_k\!\equiv\!\alpha_- \cos(k/2)$. 
For a given momentum $k$, the matrix $\mc{V}$ has the following form in the basis $\{|\psi_{k, d; S,M}^{(0)}\rangle; d=1,\cdots,d_{\text{max}}\}$:
\be\label{eq:Vprojected}
\mc{V} \!=\!\!\left(\!\!\begin{array}{cccccccc}
\beta_S & \alpha_k& 0 & \cdots &0&0&0&0\\
\alpha_k & 0 &  \alpha_k& 0 & \cdots & 0&0&0\\
0 &\alpha_k & 0 &  \alpha_k& 0 & \cdots &0&0\\
0 & 0 & \alpha_k & 0 &  \alpha_k & 0 & \cdots &0\\
\vdots&\vdots&\vdots&\vdots&\vdots&\vdots&\vdots&\vdots\\
0&0&\cdots&0&\alpha_k&0&\alpha_k&0\\
0&0&0&\cdots&0&\alpha_k&0&\alpha_k\\
0&0&0&0&\cdots&0&\alpha_k&0
\end{array}\!\!\right).~~~~
\ee

Note that, for $\alpha_-=0$ (i.e., $J_2=J_1'/2$), the matrix becomes diagonal and we get one negative eigenvalue $\beta_S<0$, the same for all $k$, with corresponding eigenstate $|\psi_{k, 1; S,M}^{(0)}\rangle$, the maximally localized bound state. This result, which is true to leading order in $J_2$ and $J_1'$, gives rise to three flat bands, corresponding to singlet bound states with energy $-\alpha_+$, triplet bound states with energy $-\alpha_+/2$, and quintuplet anti-bound states with energy $+\alpha_+/2$. 
Note also that the $k=\pi$ parts of these bands remain at the same energy irrespective of the value of $\alpha_-$, because $\alpha(k=\pi)$ vanishes. This result is again true only to leading order in $J_2$ and $J_1'$.

The energies of the two-particle states can be obtained by diagonalizing $\mc{H}$ inside the two-particle sector and using Eq.~(\ref{eq:Vprojected}). The results for a system with 24 rungs have been presented in Fig.~\ref{fig:PT}\,(a) and analysed in the main text. 


%
%

\begin{figure}[!t]
\includegraphics[width=0.99\columnwidth]{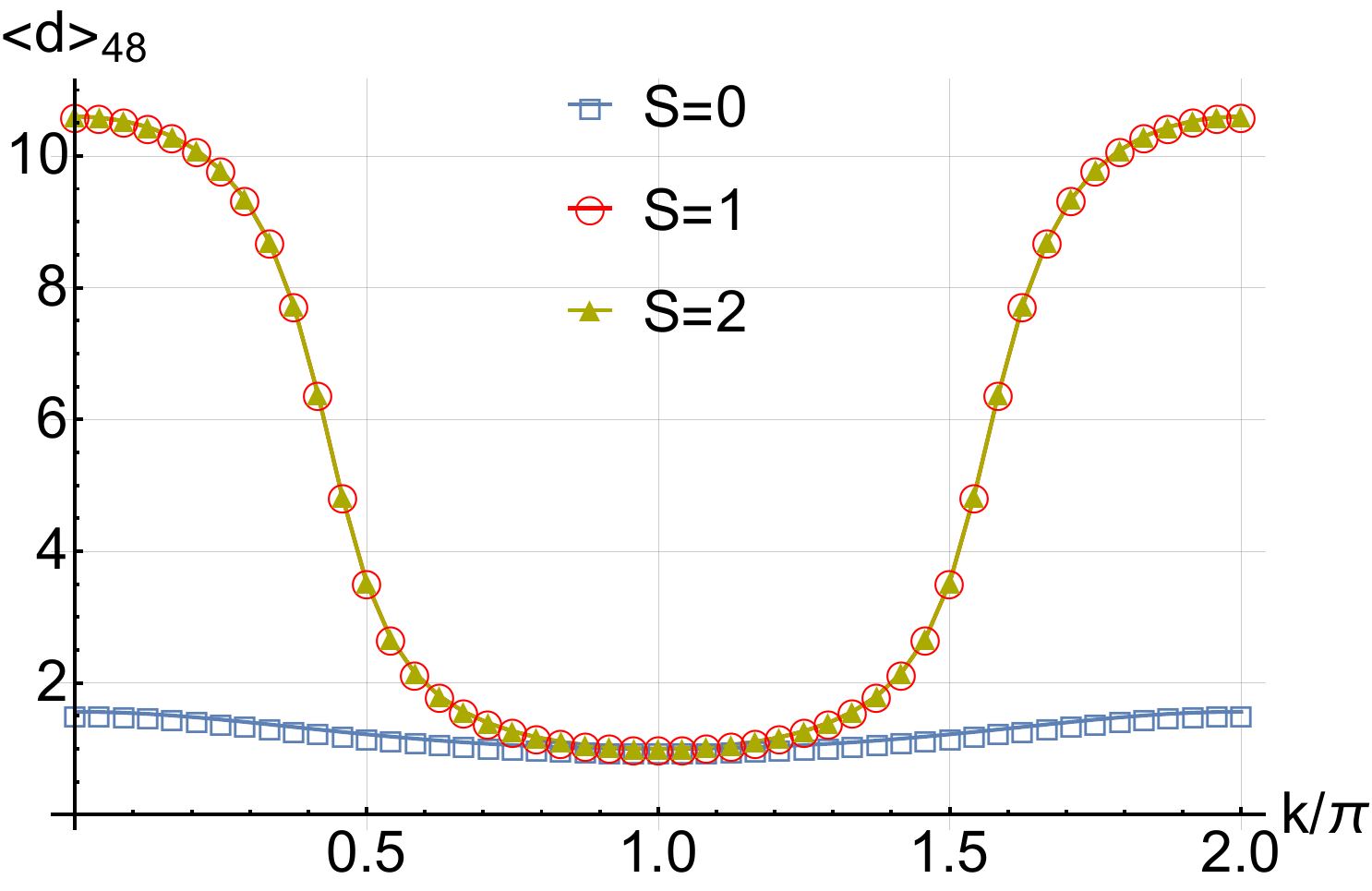}
\includegraphics[width=0.99\columnwidth]{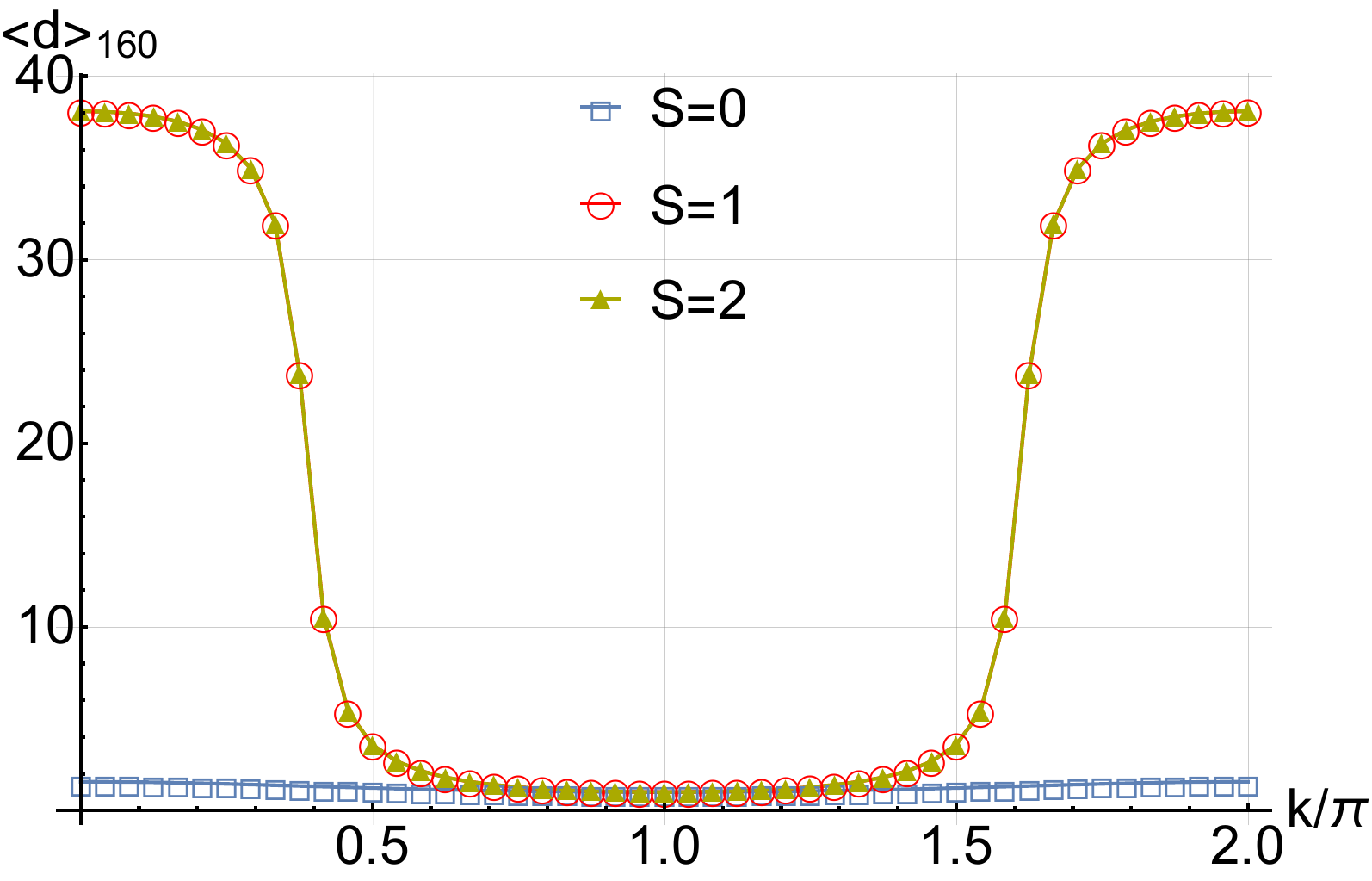}
\caption{Coherence length $\langle \hat{d}\rangle$ of the $S\!=\!0$, $S\!=\!1$ and $S\!=\!2$ bands that arise from pairs of triplets, as obtained from the corresponding eigenstates of the matrix (\ref{eq:Vprojected}) for systems with $N_r=48$ rungs (upper panel) and $N_r=160$ rungs (lower panel), using Eq.~(\ref{eq:CL}).}
\label{fig:CL}
\end{figure}

To explore the nature of the two-particle states further we examine the expectation value of the distance between the triplets, $\langle \hat{d}\rangle$, termed `coherence length' by Zheng {\it et al.}~\cite{Zheng2001}. Denoting the corresponding states by
\be
|\psi_{k;S,M}\rangle = \sum_{d=1}^{d_{\text{max}}} f_d |\psi^{(0)}_{k,d;S,M}\rangle\,,
\ee
the coherence length is defined as
\be\label{eq:CL}
\langle \hat{d}\rangle = \frac{\sum_d d~|f_d|^2}{\sum_d |f_d|^2}\,.
\ee
Figure~\ref{fig:CL} shows numerical results for the coherence length of the above $S=0$, $S=1$ and $S=2$ bands, as obtained from the corresponding eigenstates of the matrix (\ref{eq:Vprojected}) for systems with $N_r=48$ rungs (upper panel) and $N_r=160$ rungs (lower panel), using Eq.~(\ref{eq:CL}). 
We can make the following observations:

i) the coherence length of the $S\!=\!1$ and $S\!=\!2$ bands are identical. This is due to the antisymmetry $\beta_{S=2}\!=\!-\beta_{S=1}$. 
  
ii) the coherence length of all bands equals one (the minimum possible value) at $k\!=\!\pi$. This is due to the fact that $\alpha(k\!=\!\pi)$ vanishes  (see discussion above). 

iii) the coherence length of the $S\!=\!0$ band remains short throughout the momentum space (its maximum value is around 1.5 at the $\Gamma$ point) and shows very little dependence on $N_r$. This proves that the entire singlet band consists of bound states.

\begin{figure}[!b]
\includegraphics[width=0.99\columnwidth]{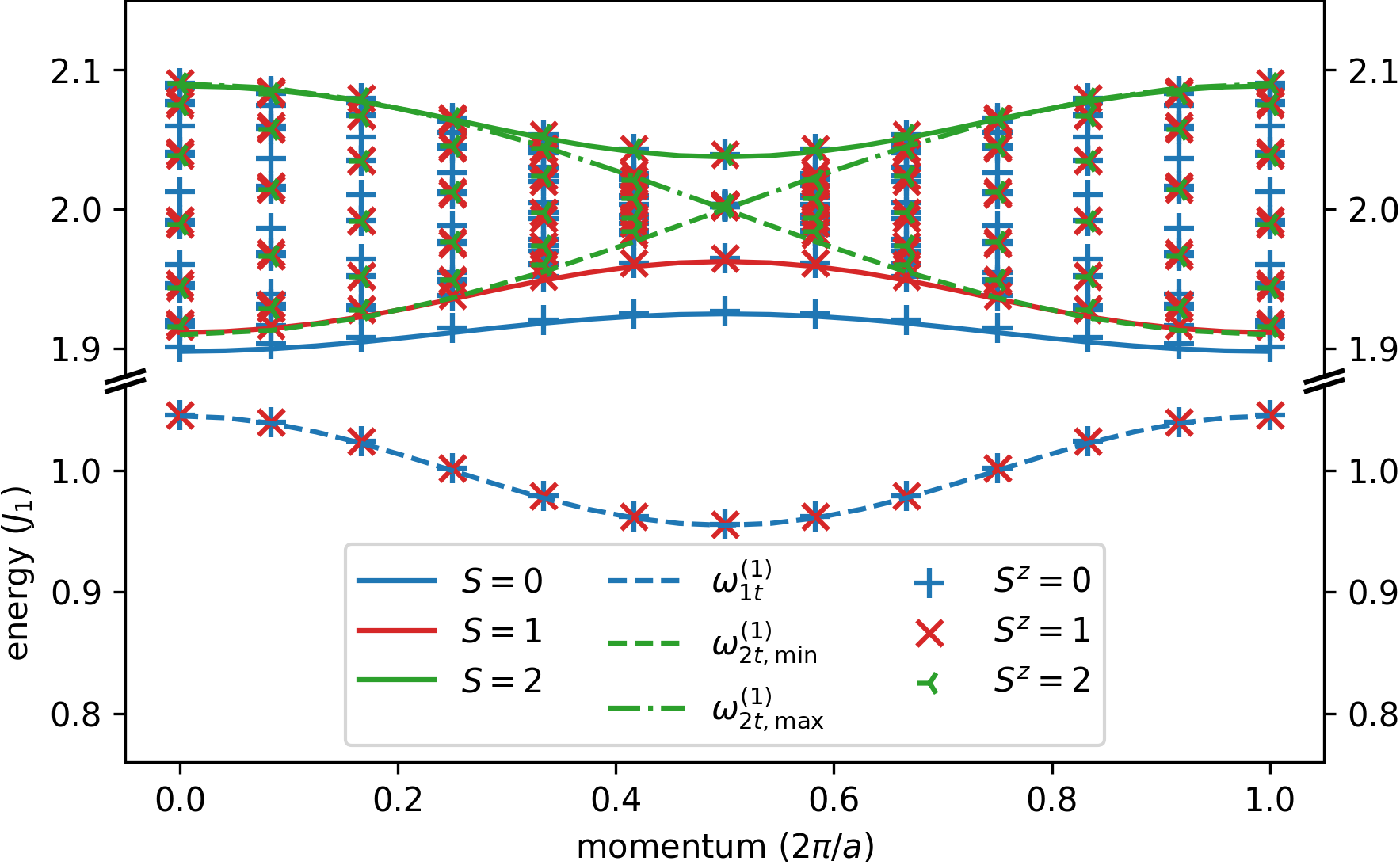}
\caption{Comparison between ED spectra of the 24-site cluster (symbols) and first-order perturbation theory predictions for the single-triplon band $\omega^{(1)}_{1t}$, the corresponding boundaries $\omega^{(1)}_{\text{2t,min}}$ and $\omega^{(1)}_{\text{2t,max}}$, of the non-interacting two-particle continuum, as well as the bound and anti-bound two-particle states (denoted by $S=0$, $S=1$ and $S=2$).}
\label{fig:ED24PT}
\end{figure}

iv) By contrast, the coherence length of the $S\!=\!1$ and $S\!=\!2$ bands is independent of $N_r$ only in a finite region around $k\!=\!\pi$. Outside this region, $\langle\hat{d}\rangle$ grows linearly with $N_r$, approaching the value $d_{\text{max}}/2$ as $k\!\to\!0$, showing that these are scattering states. The finite-size extrapolation of $1/\langle\hat{d}\rangle$ shown in Fig.~\ref{fig:PT}\,(b) of the main text confirms that there are indeed two separate momentum regimes, one where the coherence length diverges ($|k|\!\leq\!0.375\pi$; band is touching the bottom edge of the continuum), and another one with finite (and short) coherence length (states below the bottom edge of the continuum).

Finally, to validate the results of first-order perturbation theory presented above, we show in Fig.~\ref{fig:ED24PT} a comparison with ED spectra of the 24-site cluster with periodic boundary conditions for the parameters $J_1\!=\!1$, $J_1'\!=\!0.03$ and $J_2\!=\!0.06$, which is closer to the isolated rungs limit. 


%

\end{document}